# On Selfish Memes
## culture as complex adaptive system

**Hokky Situngkir**
(hokky@elka.ee.itb.ac.id, quicchote@yahoo.com)
Dept. Computational Sociology
Bandung Fe Institute

**Abstract**
We present the formal definition of meme in the sense of the equivalence between memetics and the theory of cultural evolution. From the formal definition we find that culture can be seen analytically and persuade that memetic gives important role in the exploration of sociological theory, especially in the cultural studies. We show that we are not allowed to assume meme as smallest information unit in cultural evolution in general, but it is the smallest information we use on explaining cultural evolution. We construct a computational model and do simulation in advance presenting the selfish meme power-law distributed. The simulation result shows that the contagion of meme as well as cultural evolution is a complex adaptive system. Memetics is the system and art of importing genetics to social sciences.

**Keywords:** meme, memetics, memeplex, cultural evolution, cultural unit, complex system.

*He divines remedies against injuries; he knows how to turn serious accidents to his own advantage; whatever does not kill him makes him stronger.*
F. Nietzsche, Ecce Homo (1908)

Evolution is ubiquitous. It has become the language of as many as the scientific discussions in our civilization to day. The term was come from biology, explaining how and why we have biological system as we perceive, how our species become the only creatures to which the earth count on her life. But the highest potential in the idea of evolution is not merely in biological systems. The greatest role of the theories is now explaining the dynamic of our society and culture. The paper presents a still working of evolutionary process we are going on, and how cultures become selfish in order to sustain human living above earth: memetics.

There have been a lot of discussions about the definition of memetics and how useful it is to the social analysis on explaining the social evolution (Sartika, 2004). The inventor of the term, Richard Dawkins (1976), coined the term in the motivation of seeing the cultural evolution in the sense of natural selection – a usual and sound way of thinking as the works on "culturgen" (Cavalli-Sforza, 1986), to see the cultural evolution by using the perspectives in genetics. Dawkins defined the term as the new replicator or the unit of cultural transmission or unit of imitation. In short, meme can be seen the way a cultural object or system transmitted from one person to another in the perspective of virus of mind (Brodie, 1996 & Lynch, 1998) but totally different with conventional epidemiology cases (Bartholomew, 1982:248-272), memetics not only talk about the diffusion of cultural objects or system, but also the changes occurred to the spreading object as long as the contagions.

The situation become complicated in the social science dealing with many previous concepts of culture. The paper attempts to grow memetics evolution in the theory of culture by giving any structured perspectives about culture itself and introducing memetics as a fascinating tool to comprehensively analyze socio-cultural phenomena. The next step is then to have theoretical consequences on memes and cultures, on memetic evolution to the cultural evolution. Interestingly, the paper shows power-law distributed memes in social system concerning the theme of seeing the cultural analysis as complex adaptive system in the phase of self-organized criticality.

## 1. The Cultural Evolution

Let us say, $M = \{X_1, X_2, X_3, ...\}$ as a finite (and dynamical) set of any cultural institutions of certain human society at a certain place. The cultural institutions can be certain way of life or belief systems, traditions, music genre, etc. with certain membership degree to the society. The set $M$ is realized to be dynamic since the members of the set change through time affected by the dynamic of the society – endogenously the flowering of the culture and exogenously through the cultural assimilations, acculturations, communications, and interactions of the individuals of the society with the other society.

The cultural institutions ($X_i \in M$) are then indexed in $M$ with $i$ by the sequence of the emerging cultural institutions in the society. We must note that the index $i$ is not an absolute index, but to simplify the dynamical changes of the members of $M$. The cultural institutions are also sets whose many cultural objects (abstract concepts and concrete objects) play the role in them. For example, Islam as a cultural institution is composite of cultural objects like praying-time, belief in one god, certain time for fasting, and also some concrete things like the mosque with (certain architectural design from place to place), calendar system, etc. with also some devotions or holy symbols. In other words, the cultural objects are them which are material artefacts (e.g., tools, weapons, buildings, works of art) distinctive forms of behaviour (e.g., songs, rituals, institutions, organizational forms) and system of distinctions (classifications, histories, knowledge coded in symbols, ideas or beliefs)[1]. The medium of them is signs, thought, knowledge, and language through which is passed on and transmitted from generation to generation as man's second nature (Hall, 1980). Then, we can write that the cultural objects, cultural institution, and local culture with the connection of $x_j^i \in X_i \in M$.

Important thing to note about culture is that culture itself can be viewed in many scopes of description. The term "culture" can be viewed as general term as defined above, but practically, culture is not a static variable. According to Hebdige (1987:148), culture can also be viewed as "counter-culture". Counter-culture refers to that amalgam of 'alternative' middle-class youth cultures – the hippies, the flower children, and the yippies – which grew out of the 60s, and came to prominence during the period 1967-1970. These groups are "explicitly political" and oppose dominant culture ideologically through the production of "political actions, coherent philosophies, manifestoes, etc. In the other hand, there is also term "subculture", refers to that does not necessarily directly challenge the dominant culture; rather, it can serve as a critique of the dominant culture while remaining a part of that dominant culture.

---
[1] As described by F. Heylighen (2000).

By this terminologies, we extend our definitions above on some possible ranges of $M$, as the set of cultural institutions on various sets of cultural institutions in a community ($S = \{M^A, M^B, M^C ...\}$) and the universe of our objects is the collection of every cultures there exist in a certain community. We can draw a Venn-diagram in figure 1, to ease our understanding.

Thus, the evolution of culture can be identified as the changes and transformations among any cultural objects inside certain cultural institutions ($x_m \leftrightarrow x_n$), cultural institutions ($X_p \leftrightarrow X_q$), and interactions among cultures with any other cultures from different communities ($M_1^A \leftrightarrow M_1^B, M_1^A \leftrightarrow M_1^C, M_1^B \leftrightarrow M_2^B, ...$).

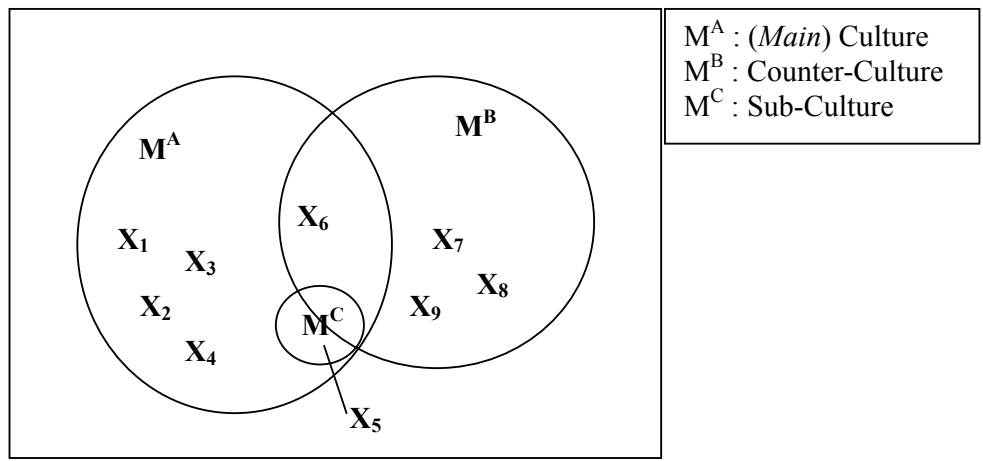

**Figure 1**
Venn diagram on culture in many scopes of description

From the description above, then we ask a question: what is the smallest unit of cultural transmission? It is certainly not easy to answer such a question, since culture is not a physical object in human nature. Culture is a collective abstraction in individual's brain[2] about her life, henceforth the endeavour to discover it in physical objects is obviously impossible. Culture should be viewed as system of symbols by which people confer significance on their own experience (Geertz, 1973). Culture must be seen as an abstract thing present to perceiving interactions and communications among individuals. In advance, we cannot also bring this out to the individual's inner cognitive system *par excellence*, since we are dealing with the **collection of cognitive systems**. In this description level, main focus is the information about how an individual perceiving her world. If language is an elementary unit of the information, specifically we are dealing with the semantics used by individuals[3]. It is all about meaning on how something presents in someone's brain in her interaction with others (man or other objects) and in return, her pay off she gains from the interaction.[4].

---

[2] We must concern that the use of the term "brain" should not be followed by the question: in what part of brain? The term is also used by Dawkins (1976) and Blackmore (1998) to emphasize that it is in a mind of an individual. Individual become the one to be the "vehicle" of the replicating memes. This is an abstraction about the abstraction of individual's mind.

## 2. Meme: Analytical Tool on Cultural Evolution

There have been a lot of definitions proposed for meme since coined for the first time by Dawkins (e.g.: Blackmore, 1998, Lynch, 1998, Wilkins, 1998) as described carefully in Sartika (2004). There are also many caveats on the existing definitions (e.g.: Gil-White, 2002) as a speculative terminologies not necessary to explain social evolution. If we are just trying to grow the usage of term meme in cultural evolution independently with the conventional cultural studies, it is apparent that the term can be a little absurd. This will **not** happen if we see meme **not** as a certain part physically existed in the process of cultural evolution, but meme as an analytical tool to approach the cultural evolution. We must remember that memetics are ontologically related to sociological analysis despite conceptually related to biology. There are some cases meme behaved like gene, but in some other part, it is not like gene at all. As noted by Heylighen (1994), our primary focus is certainly not the constitutive of a meme but to distinguish the phemetypical effects of it[5].

### 2.1. The Use of Meme in Cultural Studies

The modern history of the theoretical exploration on culture in social science is described in Situngkir (2003) that turns out to be the way to construct the description of culture in daily society's life remaining the cultural studies. The social system approached by using the computational agent-based model regards the micro-interaction of social system, in which any abstraction of the situation becomes the reference of each agent to decide her decision. In the other hand, memetics approaches social system by regarding the abstraction and its changes through the micro-interaction. The center of interest is different then, in memetics we are not dealing with the dynamics of agents' decision in micro-interaction, but the dynamics of the abstraction about the interaction that replicates in each interaction.

Before we construct any formalization about meme and how we use it to cultural studies, we should realize a technical difference with its genetic counterpart[6]. Many concepts of gene are employed in memetics, especially the process that when used in the fields of social sciences are then occupied in computation. According to Radcliffe & Surry (1994), the major difference between both is: before a meme is passed on, it is typically adapted by the person who transmits it as that person thinks, understands, and processes the meme, whereas genes get passed on whole. This character of meme is what we see relevant to understand cultural process in society.

As described in previous section, culture is multi-layer in hierarchies of description object, parts constituting the higher level of description non-linearly and so on. The important note we have from works on conventional cultural analysis is that culture is developed in the ways of how cultural units influence each other. In other words, we can say that man develops culture through "imitate" other cultural unit in her experiential space of culture. By means imitate, we refer to any different cultural unit by giving it membership degree to our own cultural space. The result is very often will not the same as the culture we copy, depends upon our own cultural system, our intelligence, and many other cognitive and external factor. The cultural unit is nonetheless is various levels of scope. It can be as simple as any melodious sound, fashion, and even as large as belief system. This proposition explains why there are many differences between the punk communities in Europe with their counterpart in Indonesia, for example, although it must be acknowledged that punk communities in Indonesia refer their culture to the ones in Europe. Therefore, we define imitation as a matter of putting some other cultural unit into our own cultural system. This is how cultural unit replicates and how culture improved *sui generis*. Culture is improved by replication of cultural unit. This is the general term of phenomena on contagious cultural aspects from places to places, including the contagion of thoughts and ideas.

---

[3] Regarding this fact, we must also incorporate meme as a semantic classification, not a syntactic one that might be directly observable in 'brain language' or natural language (Dennett. 1995:353-4).

[4] A formal model of culture regarding this can also be seen in Stahovski (1999).

[5] Like its genetic counterpart, genotype → phenotype, memetics proposed the term memetype → phemetype.

[6] This technical different is as an impact of different points between both as proposed by Sartika (2004).

As proposed by Dawkins and major memeticists, meme is the cultural unit that imitated (Blackmore, 1998) as an abstraction and neurally-stored in the brain (Lynch, 1998). Since it is an abstraction, we are not allowed to assume meme as **smallest** information unit in cultural evolution in general, but it is the **smallest** information **we use** on explaining any cultural evolution. Thus, meme can be a very small part of cultural objects (e.g.: note of music, the way use of shoe) and even the big part of culture (e.g.: nationalism, religion). In other words, meme is a matter of analytical tool on explaining culture and its dissemination, propagation, and in general, evolutionary process.

## 2.2. Meme as model for culture

Meme is a replicator unit in culture, by means of the way copies of self-imitation is made in its propagation. We understand that all of the meme constitute the memeplex that in memetic process we denote as memetype relating with phemetype.

Each meme concerns with allomeme. Allomeme, denoted as $A$, is the mutually exclusive cultural trait (Gatherer, 2002:1.2) – the alternatives meme at particular memeplex. If we have memeplex $Q$, constituted by $n$ memes, then $Q$ is the formal vector of meme values, such as

$$Q = A_1 \times A_2 \times ... \times A_n \quad (1)$$

Accordingly then, we can assume that $A$ as set of all allomemes,

$$A = \bigcup_{i=1}^{n} A_i \quad (2)$$

Practically, we concern with the relation between $Q$ as set of memetype with the set of phemetype[7] $P$ by function:

$$\psi : P \to Q \quad (3)$$

---

[7] Pheme can be defined as single memetic interactive trait which is the expression through some behavioural regularity relating to certain meme at the level of selection. In simple, pheme is the least type of selectively biased behaviour relative to the observed culture (Wilkins, 1998).

$\psi$ is injective, means that every $p \in P$ has certain unique, and well defined meme of $\psi(p) \in Q$ that represents it[8]. There is no requirement to assume that all members of $Q$ correspond to $P$.

Memetic process is then defined as the function of

$$\mu : C \to C^* \qquad (4)$$

where $\mu$ as process on maximizing the fitness of the meme as global optima $C^* \subset C$ in cultural system. Globally, we can say that $\mu$ is the function of **cultural optimization** regarding the fitness of any cultural traits to the particular social system.

In the process of its evolution, memeplex is propagated from "one brain to another", in optimally two steps (Gatherer 2002:1.3), i.e.:
1. *Horizontal transmission*, process of cultural exchange between non-familial individuals.
2. *Vertical transmission*, forms of memetic process between related individuals.

In a family, the two parents inherit memeplex to their children, which can be formalized the same as the genetic process, such as:

$$\rho : Q \times Q \times C_\rho \to Q \qquad (5)$$

where $C_\rho$ is the control parameter of the recombination process. This process can somehow followed by mutation with certain mutation rate ($C_\upsilon$, mutation control), as:

$$\upsilon : Q \times C_\upsilon \to Q \qquad (6)$$

In the other hand, replications can also happen horizontally on copying process from one person to another regarding the highest fitness value between both (controlled by $C_\sigma$), such as:

$$\sigma : Q \times Q \times C_\sigma \to Q \qquad (7)$$

This process can also followed by mutation defined in (6).

Therefore, we can say that memetic process vertically ($\mu_V = \rho \circ \upsilon$) and horizontally ($\mu_H = \sigma \circ \upsilon$) respectively yielding functional definitions:

$$\mu_V : Q \times Q \times C_\rho \times C_\upsilon \to Q \qquad (8)$$

and

$$\mu_H : Q \times Q \times C_\sigma \times C_\upsilon \to Q \qquad (9)$$

As described in Dawkins (1982), the whole process of evolution of complex adaptations requires memeplex that should follow three additional properties, i.e.: *fidelity* (the accuracy of the copying process remained by the control parameter of constants $C_\rho$, $C_\sigma$, and $C_\upsilon$), *fecundity* (the ability to generate more than one copy memeplexes in the populations), and *longevity* (the survival of memeplex, regarding concerning its fitness value).

---
[8] As adapted from Radcliffe & Surry (1994).

Memetic process is much faster than genetic process since the memetic process is highly inherent in the interactions of human being. Practically, as exemplified by simulation in the next section we can simplify the process of $\mu_H$ and $\mu_V$ by using only one parental $Q$ as done by Castro & Toro (2002). Shortly, we can say that this is the formal definition from which we are going to analyze the cultural evolution in the society.

In its memetic evolution, all of the memeplex behave to be selfish that will always spread through as many as possible human brains. It is its fitness value submitted to each of meme-configuration becomes the boundary for a meme in cultural evolution. The fitness values also shape the distribution of memeplex in population to become not uniform, and several memeplex is then dominating the whole population.

From the formal definition, we can see that culture, i.e.: the cultural institutions ($x_m \leftrightarrow x_n$), cultural institutions ($X_p \leftrightarrow X_q$), and interactions among cultures with any other cultures from different communities ($M_1^A \leftrightarrow M_1^B, M_1^A \leftrightarrow M_1^C, M_1^B \leftrightarrow M_2^B, ...$), can be represented as $Q$ interchangeably, depends upon the various measurement-level of cultural evolution we want to explain.

The hierarchical cultural system is then shape our memetic model to be also hierarchic. As pointed by Wilkins (1998), we can regard the collection of memeplex as "deme". A deme is a population of memeplex whose distinct memetic characteristics. This concept is then made it suitable to model any level of cultural unit into the language of meme. This is what we are going to do in the next section.

From this point of view, we can see that sometimes we are dealing with several demes when we process some memplexes. In other words, memeplexes are interacting with each other on emerging their deme. This is interesting as described in the section 4 and as described in Situngkir (2003), to have paradigm on seeing human culture as complex adaptive system in which the micro-macro link and emergence occurs.

## 3. Computed Culture Simulated

It is easy for us now to construct memetic analysis by using the formal definitions presented in the previous section. For the sake of simplicity, we can use the terminology proposed by Heylighen (1993) that allomeme can be formally understand as "**IF…THEN…**" propositions and submit the binary values **(0,1)** of each meme on it. The binary values may represent "yes" or "no" solution for each represented meme. In addition, we can also use the landscape of the simulation as presented by Axelrod (1997) for the culture to be observed on dissemination among agents. An exemplified algorithm used in the paper is described in figure 2.

As described in the algorithm, we can see that in each round of simulation, agents have three possible actions, *viz*
1. transmit memeplex horizontally.
2. reproduce agent with the same memeplex
3. die

All of the transmission processes are then controlled by parameters of mutation to find its highest possible fitness in its evolution.

In the simulation, we can see the evolution of each memeplex, how its population fluctuates through time. From the result of simulation we do, as figured in figure 4, we can see that the memeplex does not have any static equilibrium.

We note that the dynamics of memeplexes are representing deme in larger scale of observation. Figure 2 shows, that the fluctuation of population of each deme is constituted by certain memeplexes. In this case, we are talking about the culture not in a large scale of level description, but talking about the constituents of a cultural unit. In the level of memeplexes, we cannot see that eventually in the deme-level the sub-culture has been so strong: the population of them who has it has become very large as the main culture itself. This discussion is however frequent in the socio-cultural analysis: when the emergence of culture should be considered with the parts constituting it. The fluctuation of each meme is in figure 3 (left-side). It is obvious that in some certain length of time there can be a highly adopted memeplex in the population and in the next time it is substituted by other regime. The dominant memeplex happens to be not always become the dominant one in all of the evolutionary process. It shows us that memetics can be very useful to see how some memeplex interacts in the predefined artificial societies.

```
initialization
      generate memeplex constituted by n memes;
      generate agent with certain memeplex;
      generate randomly distributed fitness values on each memeplex;

while (halt=0)
      LOOP for the total number of agents
            choose action A(t);
            {horizontal or vertical transmission, or die}
            if (A(t) := horizontal)
                  find(neighbor)
                  compare fitness memeplex;
                  choose highest memeplex from both;
                  on probability mutate;
            end
            if (A(t) := vertical)
                  create agents with the same memeplex;
                  on probability mutate;
            end
            if (A(t) := die)
                  delete agents;
            end
      end

      evaluate population;
end
```

**Figure 2**
The exemplified algorithm for simulating culture in memetics

In research methodology, by using the memetics can be meant to see the fitness value of each memeplex regarding allomeme to be the analytical theme of research. This is a shift paradigm hypothetically proposed by the paper, that to analyze a cultural unit, we need to know the objects constituting it. The survey analysis is nonetheless needed here in order to have list of memeplex and its fitness value. The result of the survey then can be used to make artificial culture and available to be analyzed computationally. As a matter of dynamical system approached by computational method, we can also incorporate the fitness value that is changed through time. In some cases of culture, it occasionally happens that the fitness of some cultural unit varies through time.

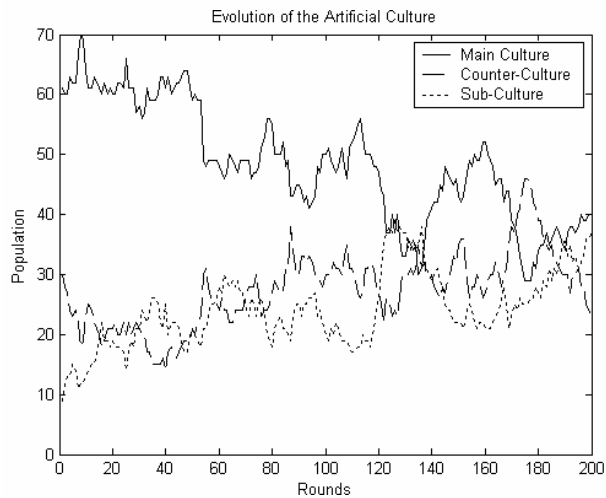

**Figure 2**
Simulation result showing the population of culture represented by demes.

Technically, in our simulation we use the static pre-defined fitness value by randomization as figured in table 1.

Table 1
Fitness value of each memeplex

| Memeplex | Fitness Value | |
|---|---|---|
| 1111 | 0.9501 | |
| 1011 | 0.2311 | Counter-culture |
| 1101 | 0.6068 | |
| 1110 | 0.4860 | |
| 0111 | 0.8913 | Sub-culture |
| 1010 | 0.7621 | |
| 1001 | 0.4565 | |
| 0101 | 0.0185 | |
| 1100 | 0.8214 | |
| 0110 | 0.4447 | |
| 0011 | 0.6154 | Main-culture |
| 1000 | 0.7919 | |
| 0100 | 0.9218 | |
| 0010 | 0.7382 | |
| 0001 | 0.1763 | |
| 0000 | 0.4057 | |

Thus, the analytical system we use here is to grow the dissemination of culture in the computational tools (Epstein & Axtell, 1996), as an artificial culture. We can see how some cultural unit evolves in the society and do some experiments on it. This method can be very useful for development and exploration on cultural phenomena. It is not too speculative if we say that memetics promise the quantitative method on analyzing culture.

In advance, it is interesting also, as described in figure 4, we can see that eventually the selfish meme is truly selfish to dominate the population adopting the fittest memeplex. This is showed by the distribution of adopted memeplex as a result of simulation. The distribution is the so-called power law distribution. We will have the theoretical elaboration about this in the next section.

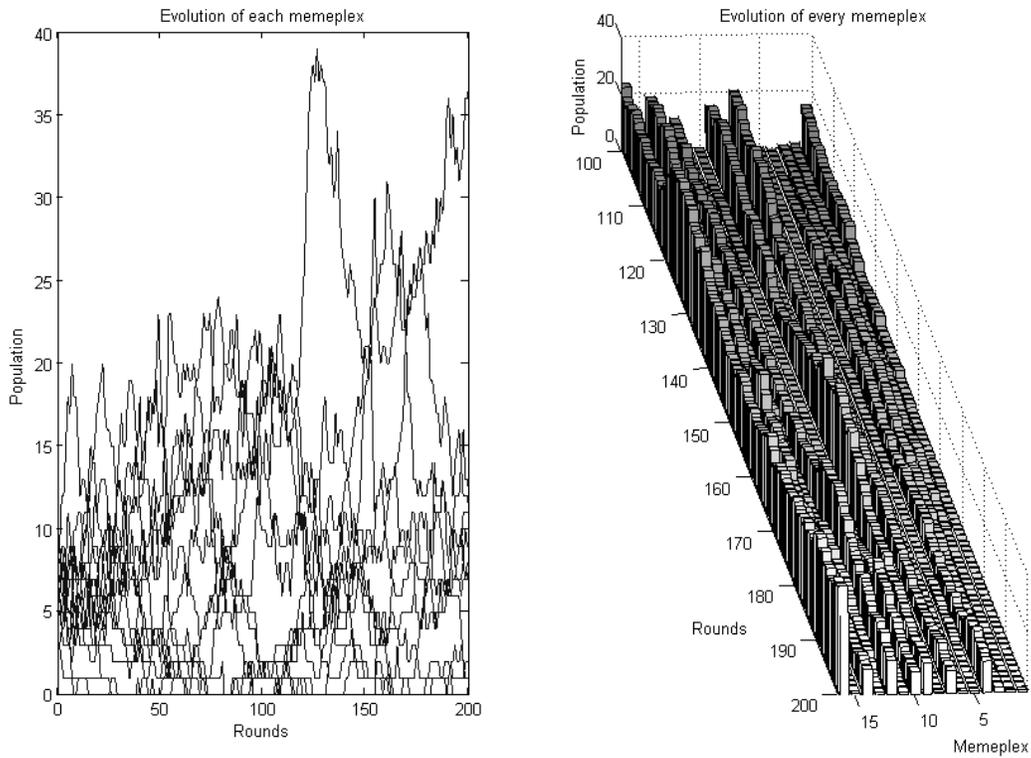

**Figure 3**
Simulation result for the evolution of memeplexes.

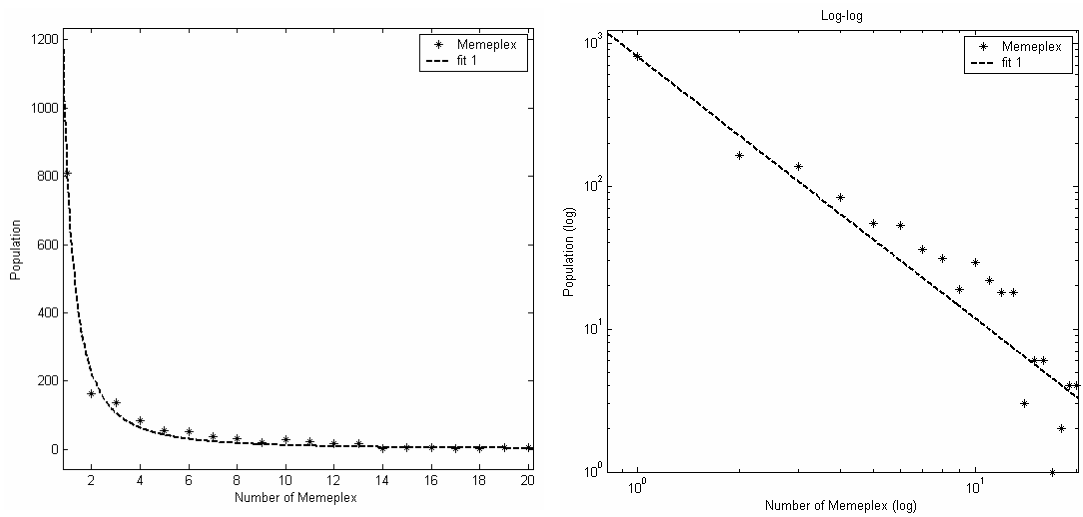

**Figure 4**
The Distribution of Memeplex in population follows the so-called power-law distribution: certain memeplexes are dominating the population of memes linearly-fitted with exponent 1.828.

## 4. Theoretical Exploration and Impacts

Imagine that we have two extreme conditions in our cultural world. The first condition is the cultural system in which there is only one fittest memeplex in the world dogmatically. The consequence is that other memeplexes will certainly disappear afterward. In the other hand, we have second condition that the memeplexes are not selected by their fitness value at all, in other words, all of the memeplexes have the same fitness value. What happens then is that the cultural process will depend on biological factor such as death-rate, birth-rate, etc. and everybody in the society closes ears and eyes and never changes memeplex. The two conditions are extreme condition, by which our real cultural evolution evolves.

In the first condition, a certain memeplex dominates, but in the second one it is showed that all of the memeplex rule simulatenously. It is trivial for us to say, that the first condition is the utmost order situation may ever occur, and contrary, the second condition is a totally disorder situation, every body in the world believe that she is the most correct one. The situations are figured in figure 5. But what we have in reality is that only some memeplexes have enough fitness value to be existed from generations to generations.

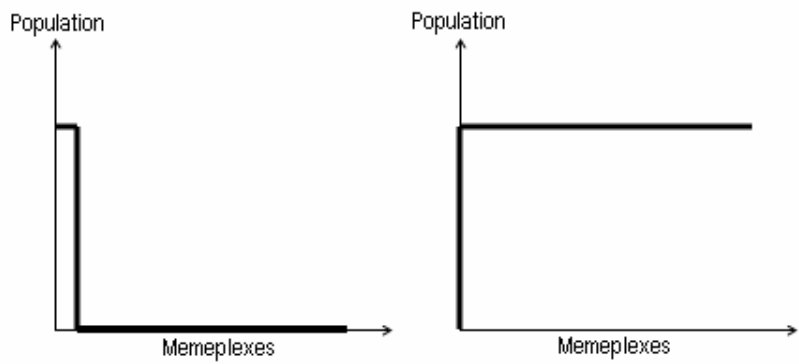

**Figure 5**
Two extreme distributions of memeplexes in the society

The two imagined cultures are presenting oversimplification to the complexity of cultural system. Yet, we realize that several memeplexes can dominate but it is obvious that the other memeplexes are not zero-distributed. Cultural evolution is laid in the critical conditions between the order and disorder, and as figured by Edmonds (1999:62), the highest complexity of cultural system is perceived around some certain points between order and disorder.

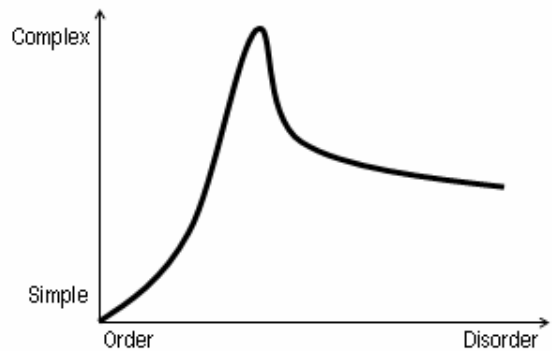

**Figure 6**
The highest complexity perceived between order and disorder (Edmonds, 1999:62)

By the experimental simulation as described in the previous section, we saw that the distribution of memeplex population is fitted with certain exponent of power-law distribution. This means that the simulating culture will always be laid around the two of extreme imagined situations. Thus, cultural patterns are emerged between order and disorder at some critical points. This is, nonetheless, become the further endeavor on analyzing cultural phenomena by using memetics. In the other hand, this signals the realization of the self-organization criticality in cultural system we have in reality: the cultural evolution is reflecting the critical points between order and fully-disorder system. A theoretical impact we have then is that in memetics, cultural system must be seen as complex adaptive system.

As a tool of cultural analysis, we can see by now that meme is a representation of diffused cultural unit. It is shown that meme concerns diffusion of the perceived; that is why memetics are close to the discussion about epidemiology of rumors (Lynch, 1998). If a meme pass through someone's brain by the process of perception, there is a process of interpretation and adoption before it goes to the next diffusion. However, the interpretation and adoption is frequent giving different output to be diffused. This is what we can see from our analysis and the above computational experiment and become the micro-properties of memetic process.

In large scale, we can say that memetics offered the multi-layer analysis of culture: the micro-macro linkage. The constituting of memeplex from memes on certain allomemes, and the existence of deme as composite of memeplexes is highly sophisticated analytical system suitable to model the highly sophisticated cultural system.

## 5. Concluding Remarks

We present here structured analysis of culture by using memetics. In memetics, we are dealing with pattern of culture as portrayed by allomeme-meme-memeplex-deme simultaneously and dynamically structured to do computational approach to it. It is also shown that memetics can be very valuable to the development of social theory by using artificial society and social simulations.

It is admitted that there are some differences of memetics terminologies presented in the paper with some other previous works, but it is to make it more structured in order to implant the memetics to the conventional cultural analysis system and taking some further advantage in the computer simulation. The experiment did in the paper opens up many possibilities on looking at cultural system in the sense of computational methods. By the structured definitions we present here, it is not about what and where actually we can see meme, whether or not meme exists in our reality, but a useful and powerful analytical tool to approach social and cultural phenomena.

Meme is however not like gene in genetics and evolutionary biology. Meme has ontological root in sociology and cultural analysis. Nonetheless, meme is the way we employ Darwinian evolution of science to the analysis of culture.

**Acknowledgement**

The author thanks Prof. Yohanes Surya for helpful financial support and colleagues in BFI for moral support. The faults remain the authors.